\documentclass[aps,prl,reprint,superscriptaddress]{revtex4-2}

\usepackage{amsmath,amssymb,amsfonts}
\usepackage{hyperref}
\usepackage{graphicx}
\usepackage{tikz}
\usepackage{braket}
\usepackage{bm}

\begin{document}

\title{Generalized Gottesman-Kitaev-Preskill States on a Quantum Torus}

\author{Sijo K. Joseph}
\email{skizhakk@gitam.edu}
\affiliation{Department of Physics, GITAM Deemed to be University, Bengaluru, Karnataka 562163, India.}

\author{Sudhir Singh}
\affiliation{Department of Mathematics, GITAM Deemed to be University, Bengaluru, Karnataka 562163, India.}
\date{\today}

\begin{abstract}
We introduce a novel formulation of a Generalized Gottesman-Kitaev-Preskill (GKP) state that resolves all of its foundational pathologies, such 
as infinite energy, non-normalizability, and orthogonality. We demonstrate that these issues are artifacts of defining the code on an unbounded phase-space. By considering the compact quantum-phase-space intrinsic to systems like coupled quantum harmonic oscillators,
we have obtained a Generalized GKP (GGKP) state that is both exact and physically realizable. This is achieved by applying an obtained Quantum Zak
Transform (QZT) to Squeezed Coherent States, which reveals that Riemann-Theta functions are the natural representation of these states on
the quantum torus phase-space. This framework not only provides a well-behaved quantum state, but also reveals a deep connection between quantum
error correction, non-commutative geometry, and the theory of generalized Theta functions. This opens a new avenue for fault-tolerant
photonic quantum computing.
\end{abstract}

\maketitle

\paragraph{Introduction--}
Gottesman-Kitaev-Preskill (GKP) code is  a cornerstone of continuous-variable (CV) 
quantum error correction, enabling fault-tolerant quantum computing~\cite{Gottesman2001}. 
Their key advantage lies in correcting small phase-space displacement errors, which is the dominant
noise in many bosonic systems~\cite{Tzitrin2020}. Despite this promise, realizing GKP codes 
experimentally remains extremely challenging. Mathematically ideal GKP states are unphysical,
they correspond to infinite superpositions of position eigenstates, are non-normalizable, and 
possess infinite energy~\cite{Gottesman2001, Jafarzadeh2025}. Consequently, all practical 
implementations rely on approximate GKP states with finite squeezing and energy, which 
suffer from reduced orthogonality and intrinsic errors even in the absence of external 
noise~\cite{Tzitrin2020, Marqversen2025}. 

The quality of these approximations depends strongly on achievable squeezing,
and the theoretical squeezing thresholds for fault tolerance lie in the range of 11.2-18.6~dB, depending on 
noise assumptions~\cite{Noh2020}, whereas current experimental platforms typically reach only 
0.62-12.91~dB~\cite{Larsen2025b,Matsos2024}. Generating approximate GKP states further requires sophisticated 
non-Gaussian operations, high-coherence storage modes, and robust stabilization techniques 
to mitigate errors~\cite{Sellem2025}.  For example, integrated photonic architectures have demonstrated on-chip 
generation of optical GKP states~\cite{Larsen2025b} but with low squeezing levels, while high-impedance superconducting circuits driven by engineered frequency combs enable dissipative protection of finite-energy GKP qubits~\cite{Sellem2025}. 
Nevertheless, all current realizations remain approximate, retaining non-orthogonality and 
limited fault-tolerance thresholds~\cite{Marqversen2025}. 

To address these fundamental issues, we develop an alternative mathematical formalism that yields 
exact, systematically generalizable GKP states. We argue that conventional GKP challenges stem from 
its formulation on an unbounded quantum optical phase-space $\mathbb{C}$. Instead, we construct the code on a compact, 
non-commutative quantum torus $\mathbb{T}_{\theta_0}^{2}$, using Rieffel-Heisenberg modules whose inner product closes naturally 
within the quantum torus phase-space algebra~\cite{ConnesRieffel1987,Rieffel1988,Jakobsen2020}, providing a well-behaved quantum phase-space description. 
Introducing a smooth 
structure on the quantum torus via its characteristics, produces regularized quasi-probability distributions, avoiding the singular 
delta comb structure of standard GKP phase-space distribution. In this framework, the usual GKP code emerges 
as a local decompactified limit of a more fundamental geometric construction. Based on these 
theoretical insights, it might be possible to generate such well-behaved GKP states 
in photonic systems, leveraging non-Gaussian resources to realize  compact-phase-space encoding.
\paragraph{Geometry of quantum-phase-space--}
Coupled quantum systems often exhibit compact underlying phase-space geometries. These geometries, including cylinders and tori, give rise to the non-commutative phase-space algebras that govern their dynamics.
A simple example of a compact quantum-phase-space is the discrete cylinder, which arises in systems with angular periodicity. 
A quantum rotor, governed by the Hamiltonian $H_{\rm rot} = \hat{L}_z^2 / (2I)$, is defined by an angular coordinate $\phi \in [0, 2\pi)$ and
 the angular momentum operator $\hat{L}_z = -i\hbar\partial_\phi$. The periodicity of $\phi$ leads to quantization of the angular momentum, 
 with eigenvalues $\ell\hbar$ for $\ell \in \mathbb{Z}$. The resulting quantum-phase-space is a discrete cylinder, $S^1 \times \mathbb{Z}$. 

In integrable classical systems with $N$ degrees of freedom, the Liouville-Arnold theorem ensures that trajectories lie on $N$-dimensional invariant tori $\mathbb{T}^N$\cite{Arnold1978}. A paradigmatic example is two coupled harmonic oscillators with some nonlinear-interaction potential $f(q_1,q_2)$ with interaction strength $\lambda$ i.e.,
\begin{equation}
H = \frac{p_1^2}{2m} + \frac{p_2^2}{2m} + \frac{1}{2}m\omega^2(q_1^2 + q_2^2) + \lambda f(q_1, q_2).
\end{equation}
The regular classical trajectories for a fixed energy lie on a 2-torus $\mathbb{T}^2 \subset \mathbb{R}^4$. Such a coupled Hamiltonian is widely explored in quantum chaos community in relation to quantum entanglement and chaos theory~\cite{skj_epjd,skj_physlett,Chung07,Chung09PRA,Lombardi11,Arul01,Zhang08,Lombardi06,Santhanam2008,Goto2021,Wang_Ent_Indi,Bandyopadhyay04,
Chaudhury09}. Upon quantization, the classical  phase-space will become a quantum torus, the periodicity of the classical angle variables motivates the use of unitary operators for modular translations in position and momentum,
\begin{equation}
U = e^{i\frac{2\pi}{L} \hat{q}}, \quad V = e^{-i\frac{2\pi}{P} \hat{p}},
\end{equation}
where $L$ and $P$ are the fundamental periods. Due to the canonical commutation relation $[\hat{q}, \hat{p}] = i\hbar$, these operators do not commute. Their algebra is given by,
\begin{equation}
UV =q V U = e^{2\pi i \theta_0} VU, \, \text{where} \, \theta_0 = \frac{2\pi\hbar}{LP} \, \text{and} \, q=e^{2\pi i \theta_0} .
\label{torus_algebra}
\end{equation}
This relation defines the algebra of the non-commutative torus $\mathbb{T}^2_{\theta_0}$~\cite{Janssen1981,Rieffel1981, Rieffel1988,Janssen1999, Connes1994,Connes1998,Manin2001,Manin2004,LuefManin2009,Jakobsen2020}. This structure is not limited to classically integrable systems, it also emerges in paradigms of quantum chaos like the quantum kicked rotor~\cite{Casati1979,Izrailev1990}, where the angular position is periodic and the momentum also becomes effectively periodic due to Floquet invariance~\cite{Santhanam2017}. Hannay and Berry even introduced a finite-dimensional quantization of quantum map on a classical torus~\cite{Hannay1980}, but here we focus on quantum system confined on a quantum torus. The quantum torus algebra will provide a unifying geometric framework for a wide range of physical systems~\cite{Connes1994}.

\begin{figure}[t]
\centering
\includegraphics[width=1.0\columnwidth]{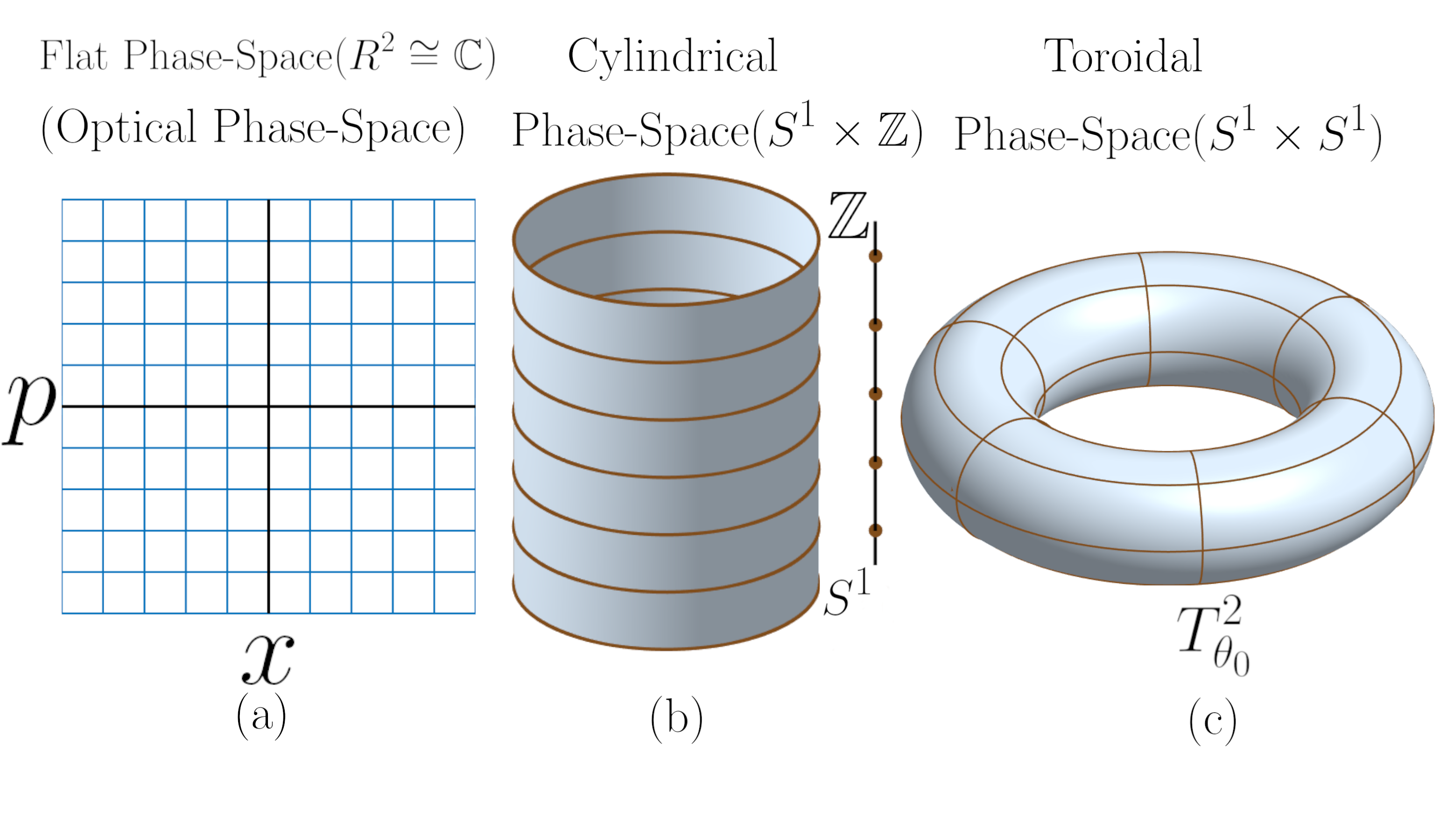}
\caption{Hierarchy of quantum phase-space geometries. (a) The unbounded phase-space plane $\mathbb{R}^2$ (or complex plane $\mathbb{C}$) of a single harmonic oscillator, where conventional GKP states are defined. (b) The discrete cylinder $S^1 \times \mathbb{Z}$ of quantum rotor models (c) The compact quantum torus (periodicity in momentum direction is imposed) $\mathbb{T}^2_{\theta_0}$ is taken as the natural setting for the Generalized GKP states proposed in this work.}
\label{phasespace}
\end{figure} 

\paragraph{Rieffel-Heisenberg Module to the Quantum Zak Transform.--}
The GKP state is built upon a lattice of displacements on flat quantum optical phase-space. The fundamental quanta of these displacements, $\alpha_0$ in position and $\beta_0$ in momentum, are determined by the periods of the torus, $L$ and $P$. We compare the modular translation operators $U = e^{i\frac{2\pi}{L} \hat{q}}$ and $V = e^{-i\frac{2\pi}{P} \hat{p}}$ with the general Weyl displacement operator $\mathcal{D}(\alpha_0, \beta_0) = \exp\left(\frac{i}{\hbar}(\beta_0\hat{q} -\alpha_0 \hat{p})\right)$. A displacement by $\alpha_0$ is generated by the operator $e^{-i\alpha_0\hat{p}/\hbar}$. Comparing $V$ with the generator, we have $ e^{-i\frac{2\pi}{P}\hat{p}} \implies {\alpha_0 = \frac{2\pi\hbar}{P}}$. This is the fundamental spacing of the lattice in the position direction. While  a displacement by $\beta_0$ is generated by $e^{i\beta_0\hat{q}/\hbar}$. Comparing this with the operator $U$, we find  $e^{i\frac{2\pi}{L}\hat{q}} \implies {\beta_0 = \frac{2\pi\hbar}{L}}$. This is the fundamental spacing of the lattice in the momentum direction. A displacement to the $(m,n)$-th lattice point is thus given by $\mathcal{D}(m\alpha_0, n\beta_0)$.
Consider the Rieffel-Heisenberg module~~\cite{Rieffel1981,Rieffel1988,Manin2001,Manin2004,Jakobsen2020}, over the smooth noncommutative torus $\mathbb{T}_{\theta_0}^{2}{}_{\infty}$, where $\theta_0$ is the deformation parameter.  In the smooth quantum torus  $\mathbb{T}_{\theta_0}^{2}{}_{\infty} \, (\mathbb{T}_{\theta_0}^{2}{}_{\infty} \subset \mathbb{T}_{\theta_0}^{2})$
 all phase-space Fourier coefficients are rapidly decaying, a natural setting for lattice sums.

As a Hilbert module we take the Schwartz space $\mathcal{S}(\mathbb{R})$,  while the physical Hilbert space of states is obtained by completing $\mathcal{S}(\mathbb{R})$ in the $L^2$-norm, yielding $L^2(\mathbb{R})$. The Schwartz space $\mathcal{S}(\mathbb{R})$ serves as the Hilbert module, with a right action $\mathbb{T}_{\theta_0}^{2}{}_{\infty}$  defined through the Weyl displacement operators $\mathcal{D}(m\alpha_0, n\beta_0)=\mathcal{D}_{m,n}$, which satisfy the twisted multiplication rule,
\begin{eqnarray}
 \mathcal{D}_{m,n} \mathcal{D}_{m',n'} = e^{i \pi  \theta_0 (nm'-m n' )} \mathcal{D}_{m+m',n+n'} 
\end{eqnarray}

According to ~\cite{Manin2001,Manin2004,LuefManin2009,Janssen1981}, $\mathbb{T}^2_{\theta_0}{}_{\infty}$ valued module inner product is given by,
\begin{equation}
\langle \phi, \psi \rangle_{\mathbb{T}^2_{\theta_0}{}_{\infty}} = \sum_{m,n \in \mathbb{Z}} 
\langle \phi | \mathcal{D}(m\alpha_0, n\beta_0) | \psi \rangle \, U^n V^m, \label{innerpro}
\end{equation}

\begin{equation}
\langle \phi, \psi \rangle_{\mathbb{T}^2_{\theta_0}{}_{\infty}} = \sum_{m,n \in \mathbb{Z}} 
\langle \phi | \mathcal{D}(m\alpha_0, n\beta_0) | \psi \rangle \, U^n V^m, \label{innerpro}
\end{equation}
where $U =  \mathcal{D}(0,\beta_0)$ and $V =  \mathcal{D}(\alpha_0,0)$ generate the quantum torus algebra given in Eq.~\ref{torus_algebra}. 
Note that, the inner product described here is an algebra valued in $\mathbb{T}^2_{\theta_0}{}_{\infty}$ (rather than in \(\mathbb C\)), as required by the module structure. Scope of such an inner product is already mentioned in the quantum-mechanical context~\cite{Moore2017} but not yet in the quantum optics point of view.
If we replace the algebra generators with continuous torus characters such that, 
\begin{eqnarray}
{\chi}_{m,0}=e^{-2\pi i m \alpha_0 k},  \quad {\chi}_{0,n}=e^{2\pi i n \beta_0 x}. \label{tcharacter}
\end{eqnarray} 
Then defining Moyal product~\cite{Groenewold1946,Moyal1949,DAscanio2016} between torus characters, one can find that ${\chi}_{0,n}\star_{\theta_0}{\chi}_{m,0}={e}^{i\pi \theta_0 m n}\chi_{m,n}(x,k)$ and it obeys all the properties of Weyl displacement operator $\mathcal{D}(\alpha_0, \beta_0)$ and therefore recovers functions on continuous quantum torus.

For a state $\psi \in \mathcal{S}(\mathbb{R})$ and a probe $\phi \in \mathcal{S}(\mathbb{R})$, taking 
torus characters given in Eq.~\ref{tcharacter} and substituting for the generators $U$ and $V$ in Eq.~\ref{innerpro}, the Rieffel-Heisenberg module inner product can be expressed as a function on the continous torus,
\begin{eqnarray}
\langle \phi, \psi \rangle_{\mathbb{T}^2_{\theta_0}{}_{\infty}}
=\sum_{m,n \in \mathbb{Z}} \langle \phi |\mathcal{D}(m\alpha_0, n\beta_0)|\psi \rangle \, 
{e}^{2\pi i (n \beta_0 x- m \alpha_0 k)}. \nonumber \\ \label{zak_general}
\end{eqnarray}
This construction naturally maps the Rieffel - Heisenberg module inner product to a quasi-probability phase-space representation on a continuous quantum torus phase-space. We call this phase-space probability distribution as Quantum Zak Transform (QZT),  much more generalized version of the Classical Zak Transform (CZT) on a quantum torus~\cite{Zak1964}.  Classical Zak Transform is recently explored in connection with GKP states~\cite{Pantaleoni2023} but here it emerges in a much more generalized form from a rich mathematical structure. QZT proposed here realizes a continous  torus function while faithfully encoding the underlying quantum torus structure. Surprisingly, this exact extended  Zak Transform  already matches with the proposed form in the mathematical literature~\cite{Zayed1995}. Here  in quantum optics, it appears as a quasi-probability distribution on a quantum torus phase space. Crucially, the Zak coefficients includes the deformation parameter $\theta_0$ that correctly models the non-commutative geometry inherent to quantum mechanical position and momentum coordinates.

\paragraph{QZT for Squeezed Coherent States--}
Consider two displaced squeezed coherent states $\ket{\psi}$ (signal) and $\ket{\phi}$ (probe),
\begin{equation}
\ket{\psi} = \mathcal{D}(\alpha_\psi)\mathcal{S}(\zeta_\psi)\ket{0}, \quad  \ket{\phi} = \mathcal{D}(\alpha_\phi)\mathcal{S}(\zeta_\phi)\ket{0},
\end{equation}
where $\mathcal{D}(\alpha) = \exp(\alpha\hat{a}^\dagger - \alpha^*\hat{a})$ is the displacement operator, and $\mathcal{S}(\zeta) = \exp\left[\frac{1}{2}(\zeta^*\hat{a}^2 - \zeta\hat{a}^{\dagger 2})\right]$ is the squeezing operator with $\zeta = re^{i\theta}$.
For displaced squeezed states, the matrix element $\braket{\phi | \mathcal{D}(m\alpha_0,n \beta_0) | \psi}$ is a Gaussian function. 
The matrix element $\braket{\phi | \mathcal{D}(m,n) | \psi}$ is computed by evaluating the Gaussian integral over position space wavefunctions. The wavefunctions for displaced squeezed states are (See Refs.~\cite{JagdishRai88,Moller96}),
\begin{align}
\psi(q) &= \frac{1}{(\pi\sigma_\psi^2)^{1/4}} \exp\left[ -\frac{(q - q_\psi)^2}{2\sigma_\psi^2} + \frac{i}{\hbar} p_\psi \left(q - \frac{q_\psi}{2}\right) \right], \\
\phi(q) &= \frac{1}{(\pi\sigma_\phi^2)^{1/4}} \exp\left[ -\frac{(q - q_\phi)^2}{2\sigma_\phi^2} - \frac{i}{\hbar} p_\phi \left(q - \frac{q_\phi}{2}\right) \right].
\end{align}
In position basis representation, the phase-space displacement operator $\mathcal{D}(m\alpha_0,n\beta_0)$ acts as,
\begin{equation}
\mathcal{D}(m\alpha_0,n\beta_0) \psi(q) = e^{i\beta_n (q - \alpha_m/2)/\hbar} \psi(q - \alpha_m),
\end{equation}
where $\alpha_m = m\alpha_0 = \frac{2\pi\hbar m}{P}$ and $\beta_n = n\beta_0 = \frac{2\pi\hbar n}{L}$. The matrix element becomes,
\begin{equation}
\braket{\phi | \mathcal{D}(m\alpha_0,n\beta_0) | \psi} = \int \phi^*(q) e^{i\beta_n (q - \alpha_m/2)/\hbar} \psi(q - \alpha_m) dq.\label{gaussintegral}
\end{equation}
Equation~\ref{gaussintegral} is a Gaussian intergal and it evaluates into Eq.~\ref{melement} after completing the square and simplifying,
\begin{equation}
\braket{\phi | \mathcal{D}(m\alpha_0,n\beta_0)| \psi} = \mathcal{N} {e}^{\Phi}\exp{\left[ -\pi \bm{m}^T \bm{\Gamma} \mathbf{m} + \boldsymbol{\eta}^T \bm{m} \right]},
\label{melement}
\end{equation}
where $\bm{m} =(m,n)^T$,  $\bm{z}=(-\alpha_0 k, \beta_0 x)^T$, $\Delta q=q_{\psi}-q_{\phi}$,
$\Delta p = p_\psi-p_\phi$, $\mathcal{N}$ is a normalization constant where,
\begin{align}
\bm{\Gamma} &= \begin{pmatrix}
\frac{{\theta_0}^2 L^2}{2\pi \Sigma^2} & \frac{i\theta_0}{2\Sigma^2} ({\sigma_\phi}^2-{\sigma_\psi}^2) \\
 \frac{i\theta_0}{2\Sigma^2} ({\sigma_\phi}^2-{\sigma_\psi}^2) & \frac{{\theta_0}^2 P^2}{2\pi \hbar^2\Sigma^2} ({\sigma_\phi}^2{\sigma_\psi}^2)
\end{pmatrix}, \\
\boldsymbol{\eta} &= \begin{pmatrix}
\eta_m \\ \eta_n
\end{pmatrix} =
\begin{pmatrix}
\frac{2\pi}{P\Sigma^2}\left[ -\hbar\Delta q - i(p_\phi\sigma_\phi^2 + p_\psi\sigma_\psi^2) \right] \\
\frac{2\pi}{L\Sigma^2}\left[ i\hbar(q_\phi\sigma_\psi^2 + q_\psi\sigma_\phi^2) + \sigma_\phi^2\sigma_\psi^2 \Delta p \right] \\
\end{pmatrix}.
\end{align}
An exponential function with phase $\Phi$ is obtained as an external factor,
\begin{eqnarray}
\Phi=\exp{\Bigl(C_{0}-\frac{B_{0}^2}{4A}\Bigr)}
\end{eqnarray}
where  $C_{0}=-\frac{q_\phi^2}{2\sigma_\phi^2}-\frac{q_\psi^2}{2\sigma_\psi^2}+\frac{i}{2\hbar}(p_\phi q_\phi-p_\psi q_\psi)$, $B_{0}={\frac{q_\phi}{\sigma_\phi^2}+\frac{q_\psi}{\sigma_\psi^2}+\frac{i}{\hbar}(p_\psi-p_\phi)}$ and $A=-\frac{1}{2}\left(\sigma_\phi^{-2}+\sigma_\psi^{-2}\right)$.
Here, $\sigma_\phi$ and $\sigma_\psi$ are the squeezing parameters, $(q_\phi, p_\phi)$ and $(q_\psi, p_\psi)$ are the phase-space coordinates of each states,  quantum deformation parameter $\theta_0=\frac{2\pi\hbar}{LP}$ and $\Sigma^2 = \sigma_\phi^2 + \sigma_\psi^2$.
Substituting Eq.~\ref{melement} into Eq.~\ref{zak_general} and rearranging terms, we obtain,
\begin{eqnarray}
& \langle \phi, \psi \rangle_{\mathbb{T}^2_{\theta_0}{}_{\infty}}  = \mathcal{N} e^{\Phi} \Theta(\bm{\xi}\mid \mathbf{\Omega}),
\label{thetaeq}\\
&\Theta(\bm{\xi}\mid \mathbf{\Omega})=\sum_{n\in\mathbb{Z}^{2}} \exp\!\Big(\pi i\, \bm{m}^{\mathsf T}\mathbf{\Omega}\, \bm{m} + 2\pi i\, \bm{m}^{\mathsf T} \bm{\xi}\Big),
\end{eqnarray}
where $\Theta(\bm{\xi}\mid \mathbf{\Omega})$ is the Genus-2 Riemann Theta function~\cite{MumfordBook1,Bernard2004}. The period matrix $\mathbf{\Omega}$ and argument $\bm{\xi}$ is given by,
\begin{align}
\mathbf{\Omega} &= i \bm{\Gamma}\\
\mathbf{\bm{\xi}} &= \begin{pmatrix} -{\alpha}_{0} k \\ \beta_{0} x \end{pmatrix} + \frac{1}{2\pi i}\begin{pmatrix} {\eta}_{m} \\ {\eta}_{n} \end{pmatrix}
\end{align}
with $\theta_0 = \frac{2\pi\hbar}{LP}$. The imaginary part of $\mathbf{\Omega}$ is positive-definite,
\begin{equation}
\Im(\mathbf{\Omega}) = \begin{pmatrix} \frac{{\theta_0}^2 L^2}{2\pi \Sigma^2} &  0 \\
0 & \frac{{\theta_0}^2 P^2}{2\pi \hbar^2\Sigma^2} ({\sigma_\phi}^2{\sigma_\psi}^2)
\end{pmatrix},
\end{equation}
ensuring absolute convergence of the theta series. Note that $\mathbf{\Omega}$, arises as a result of quantum geometric parameter $\theta_0$ and this is bringing all the regularization terms in the Fourier sum. Once $\theta_0$ is zero, all the regularization terms disappears.
This result establishes an exact correspondence between the physical parameters of displaced squeezed states and the structure of the Riemann Theta function, providing a mathematical foundation for analyzing quantum optical states  on a  compact toroidal quantum phase-space lattices. In short, non-commutativity of the quantum-phase-space bring Gaussian regularization terms. 
\paragraph{Squeezed Vacuum States--} 
When $\ket{\psi} = \ket{\phi} = \mathcal{S}(r)\ket{0}$ (Equally squeezed vacuum state with $\sigma_\phi=\sigma_\psi=\sigma$), the period matrix and argument simplify to, $$\mathbf{\Omega} = \begin{pmatrix} \tau_1 & 0 \\ 0 & \tau_2 \end{pmatrix}, \quad
\bm{\xi} = \begin{pmatrix} -\alpha_0 k\\ \beta_0 x \end{pmatrix}.$$

The cross-terms vanish ($\Gamma_{12} = 0$) due to phase alignment ( $\sigma_\phi=\sigma_\psi=\sigma$), then the  Riemann Theta function factorizes as,
\begin{equation}
\Theta\left( \bm{\xi}\mid \mathbf{\Omega} \right) = \vartheta_3\left( \xi_1,\tau_1  \right) \vartheta_3\left( \xi_2, \tau_{2} \right),
\label{factorizedtheta}
\end{equation}
where $\vartheta_3(\xi,\tau)$ is the Jacobi Theta function $ \left(\vartheta_3(z, \tau)  = \sum_{n=-\infty}^{\infty} \exp(i\pi\tau n^2 + 2\pi i \xi n) \right)$~\cite{MumfordBook1,Abramowitz1970}. Here the decay parameters are now simply,
\begin{equation}
\tau_1 = i\frac{\pi\hbar^2}{P^2\sigma^2} = i\frac{\theta_0^2}{4\pi\sigma^2}L^2 \quad \text{and} \quad \tau_2 = i\frac{\pi\sigma^2}{L^2}=\frac{\theta_0^2}{4\pi\hbar^2}\sigma^2 P^2.
\end{equation}
The factorization of the distribution in Eq.~\ref{factorizedtheta}, demonstrates the separation of position and momentum dynamics on the commutative torus. Also note that the spatial part of the distribution given in Eq.~\ref{factorizedtheta} correctly produce the coherent state of quantum particle on a circle~\cite{Kowalski1996,Kowalski2021}. The parameters $\tau_x$ and $\tau_k$ directly encode the quantum squeezing. Their product, $|\tau_1| |\tau_2| = (\pi\hbar/PL)^2 = (\theta_0/2)^2$, is independent of squeezing and reveals a fundamental lattice uncertainty principle intrinsic to the quantum-toroidal geometry. In order to move from a compactified torus to a non-compact phase-space the $L$ and $P$ parameters should be large, that will take us to a flat phase-space. It can be seen that, removing both cutoffs ($P \to \infty$, $L \to \infty$) recovers the ideal GKP grid state on flat phase-space,  
\begin{equation}
\langle \phi, \psi \rangle_{\mathbb{T}^2_{\theta_0}{}_{\infty}} \propto \sum_{m \in \mathbb{Z}} \delta(q - mL) \sum_{n \in \mathbb{Z}} \delta(p - nP).
\end{equation}
Hence, Riemann Theta function $\Theta\left( \bm{\xi}\mid\mathbf{\Omega} \right) $ provides a normalizable, finite-energy regularized GKP state on the quantum torus phase-space. Note that infinite energy and non-normalization problem is already solved via the phase-space compactification. We will show that, representing GKP state as inner product on Rieffel-Heisenberg modules further ensure algebraic consistency and enhanced orthogonality (algebraic orthogonality) compared to flat-space constructions.
\paragraph{Orthogonality of the logical states --}
Consider a vaccum reference state $|0\rangle$ and  two qumode states 
$|\psi\rangle$ and $|\phi\rangle$. All these states are already in $\mathcal{S}(\mathbb{R})$, hence $\langle\phi|\psi\rangle_{L^2}$ will never becomes zero (orthogonal) unless separated with infinite distance. This is the main technological bottle-neck in squeezed state based continuous variable photonic quantum computing. One of the available solutions is  the extreme level of quantum squeezing ($18.6dB$ theoretically desired for practical fault tolerence~\cite{Noh2020}) to make it nearly orthogonal. Mathematically it impossible to achieve the orthogonality unless infinite squeezing is applied which is practically an impossible task. 

Taking an alternate point of view in the Rieffel-Heisenberg module framework, consider two module valued inner products,
$\bm{a}=\langle 0| \psi \rangle_{\mathbb{T}^2_{\theta_0}{}_{\infty}}$ and  $\bm{b}=\langle 0| \phi \rangle_{\mathbb{T}^2_{\theta_0}{}_{\infty}}$ respectively.  These module valued inner products $\bm{a}$ and $\bm{b}$ are not in $\mathbb{C}$, they are the elements of the quantum torus phase-space algebra ${\mathbb{T}^2_{\theta_0}{}_{\infty}} $.  In the quantum optics point of view, these quantities can be further perceived as the quantum optics phase-space distributions representing coherent states on a circle with momentum compactification, which is found to be Riemann Theta function of genus-2.  Expressing Riemann Theta function in terms of its characteristic $ \bm{\epsilon}$ and   $\bm{\delta}$, it will allow a projection of the function into two sets of  even parity  and odd parity states i.e., 
\begin{eqnarray}
 \bm{a}=\Theta\begin{bmatrix} 0 \\ \bm{0} \end{bmatrix}(\bm{\xi}\mid\mathbf{\Omega}) \quad \text{and} \quad 
\bm{b}=\Theta\begin{bmatrix} {1}/{2} \\ {1}/{2} \end{bmatrix}(\bm{\xi}\mid\mathbf{\Omega}), 
\end{eqnarray} 
respectively. These are two distinct parity class of the Riemann Theta functions which are obtained as the elements of the quantum torus algebra and it can be shown that  the inner product between different parity classes in the Hilbert sense is vanishing. 
Hence one can take $\bm{a}\equiv|0\rangle_{L}\equiv|GGKP\rangle_{0}$ and $\bm{b}\equiv|1\rangle_{L}\equiv|GGKP\rangle_{1}$ as two logical Generalized-GKP (GGKP) states living on a quantum torus phase-space. At the same time, the coherent state overlap $\langle \psi| \phi \rangle_{L^2}$ can still be nonzero since it comes from canonical trace of the module inner product .
\paragraph{Conclusion--}
Inner product in Rieffel-Heisenberg Module can be seen as a quantum optical quasi-probability distribution on a quantum torus phase-space in the continuum limit.  Module valued inner product is mapped on to a continuous quantum-torus via torus characteristics. QZT is obtained in the continuous quantum torus limit which contains a natural and  unified regularization of GKP states, resolving the pathologies of non-normalizability, infinite energy, and imperfect orthogonality. Using QZT, we obtained GGKP state as Riemann Theta function which contains an  intrinsic Gaussian regulator that ensures well-defined, finite-energy, and algebraically orthogonal states. The obtained GGKP state matches to GKP state in the flat phase-space (quantum mechanical) regime. Moreover, the required toroidal phase-space geometry may emerge naturally in coherent state interactions offering clear experimental  pathways. Thus, the well behaved Generalized GKP code (GGKP) lives on quantum torus phase-space geometry  establishes a robust foundation for future photonic quantum error correcting code space.


\begin{thebibliography}{51}%
\makeatletter
\providecommand \@ifxundefined [1]{%
 \@ifx{#1\undefined}
}%
\providecommand \@ifnum [1]{%
 \ifnum #1\expandafter \@firstoftwo
 \else \expandafter \@secondoftwo
 \fi
}%
\providecommand \@ifx [1]{%
 \ifx #1\expandafter \@firstoftwo
 \else \expandafter \@secondoftwo
 \fi
}%
\providecommand \natexlab [1]{#1}%
\providecommand \enquote  [1]{``#1''}%
\providecommand \bibnamefont  [1]{#1}%
\providecommand \bibfnamefont [1]{#1}%
\providecommand \citenamefont [1]{#1}%
\providecommand \href@noop [0]{\@secondoftwo}%
\providecommand \href [0]{\begingroup \@sanitize@url \@href}%
\providecommand \@href[1]{\@@startlink{#1}\@@href}%
\providecommand \@@href[1]{\endgroup#1\@@endlink}%
\providecommand \@sanitize@url [0]{\catcode `\\12\catcode `\$12\catcode
  `\&12\catcode `\#12\catcode `\^12\catcode `\_12\catcode `\%12\relax}%
\providecommand \@@startlink[1]{}%
\providecommand \@@endlink[0]{}%
\providecommand \url  [0]{\begingroup\@sanitize@url \@url }%
\providecommand \@url [1]{\endgroup\@href {#1}{\urlprefix }}%
\providecommand \urlprefix  [0]{URL }%
\providecommand \Eprint [0]{\href }%
\providecommand \doibase [0]{https://doi.org/}%
\providecommand \selectlanguage [0]{\@gobble}%
\providecommand \bibinfo  [0]{\@secondoftwo}%
\providecommand \bibfield  [0]{\@secondoftwo}%
\providecommand \translation [1]{[#1]}%
\providecommand \BibitemOpen [0]{}%
\providecommand \bibitemStop [0]{}%
\providecommand \bibitemNoStop [0]{.\EOS\space}%
\providecommand \EOS [0]{\spacefactor3000\relax}%
\providecommand \BibitemShut  [1]{\csname bibitem#1\endcsname}%
\let\auto@bib@innerbib\@empty
\bibitem [{\citenamefont {Gottesman}\ \emph {et~al.}(2001)\citenamefont
  {Gottesman}, \citenamefont {Kitaev},\ and\ \citenamefont
  {Preskill}}]{Gottesman2001}%
  \BibitemOpen
  \bibfield  {author} {\bibinfo {author} {\bibfnamefont {D.}~\bibnamefont
  {Gottesman}}, \bibinfo {author} {\bibfnamefont {A.}~\bibnamefont {Kitaev}},\
  and\ \bibinfo {author} {\bibfnamefont {J.}~\bibnamefont {Preskill}},\ }\href
  {https://doi.org/10.1103/PhysRevA.64.012310} {\bibfield  {journal} {\bibinfo
  {journal} {Phys. Rev. A}\ }\textbf {\bibinfo {volume} {64}},\ \bibinfo
  {pages} {012310} (\bibinfo {year} {2001})}\BibitemShut {NoStop}%
\bibitem [{\citenamefont {Tzitrin}\ \emph {et~al.}(2020)\citenamefont
  {Tzitrin}, \citenamefont {Bourassa}, \citenamefont {Menicucci},\ and\
  \citenamefont {Sabapathy}}]{Tzitrin2020}%
  \BibitemOpen
  \bibfield  {author} {\bibinfo {author} {\bibfnamefont {I.}~\bibnamefont
  {Tzitrin}}, \bibinfo {author} {\bibfnamefont {J.~E.}\ \bibnamefont
  {Bourassa}}, \bibinfo {author} {\bibfnamefont {N.~C.}\ \bibnamefont
  {Menicucci}},\ and\ \bibinfo {author} {\bibfnamefont {K.~K.}\ \bibnamefont
  {Sabapathy}},\ }\href {https://doi.org/10.1103/PhysRevA.101.032315}
  {\bibfield  {journal} {\bibinfo  {journal} {Phys. Rev. A}\ }\textbf {\bibinfo
  {volume} {101}},\ \bibinfo {pages} {032315} (\bibinfo {year}
  {2020})}\BibitemShut {NoStop}%
\bibitem [{\citenamefont {Jafarzadeh}\ \emph {et~al.}(2025)\citenamefont
  {Jafarzadeh}, \citenamefont {Conrad}, \citenamefont {Alexander},\ and\
  \citenamefont {Baragiola}}]{Jafarzadeh2025}%
  \BibitemOpen
  \bibfield  {author} {\bibinfo {author} {\bibfnamefont {M.}~\bibnamefont
  {Jafarzadeh}}, \bibinfo {author} {\bibfnamefont {J.}~\bibnamefont {Conrad}},
  \bibinfo {author} {\bibfnamefont {R.~N.}\ \bibnamefont {Alexander}},\ and\
  \bibinfo {author} {\bibfnamefont {B.~Q.}\ \bibnamefont {Baragiola}},\
  }\href@noop {} {\bibfield  {journal} {\bibinfo  {journal} {arXiv:2504.13383
  [quant-ph]}\ } (\bibinfo {year} {2025})}\BibitemShut {NoStop}%
\bibitem [{\citenamefont {Marqversen}\ \emph {et~al.}(2025)\citenamefont
  {Marqversen}, \citenamefont {Michelsen}, \citenamefont {Wesenberg},\ and\
  \citenamefont {Zinner}}]{Marqversen2025}%
  \BibitemOpen
  \bibfield  {author} {\bibinfo {author} {\bibfnamefont {F.~K.}\ \bibnamefont
  {Marqversen}}, \bibinfo {author} {\bibfnamefont {A.~B.}\ \bibnamefont
  {Michelsen}}, \bibinfo {author} {\bibfnamefont {J.~H.}\ \bibnamefont
  {Wesenberg}},\ and\ \bibinfo {author} {\bibfnamefont {N.~T.}\ \bibnamefont
  {Zinner}},\ }\href@noop {} {\bibfield  {journal} {\bibinfo  {journal}
  {arXiv:2507.15955 [quant-ph]}\ } (\bibinfo {year} {2025})}\BibitemShut
  {NoStop}%
\bibitem [{\citenamefont {Noh}\ and\ \citenamefont
  {Chamberland}(2020)}]{Noh2020}%
  \BibitemOpen
  \bibfield  {author} {\bibinfo {author} {\bibfnamefont {K.}~\bibnamefont
  {Noh}}\ and\ \bibinfo {author} {\bibfnamefont {C.}~\bibnamefont
  {Chamberland}},\ }\href {https://doi.org/10.1103/PhysRevA.101.012316}
  {\bibfield  {journal} {\bibinfo  {journal} {Phys. Rev. A}\ }\textbf {\bibinfo
  {volume} {101}},\ \bibinfo {pages} {012316} (\bibinfo {year}
  {2020})}\BibitemShut {NoStop}%
\bibitem [{\citenamefont {Larsen}\ \emph {et~al.}(2025)\citenamefont {Larsen},
  \citenamefont {Bourassa}, \citenamefont {Kocsis},\ and\ \citenamefont
  {et~al.}}]{Larsen2025b}%
  \BibitemOpen
  \bibfield  {author} {\bibinfo {author} {\bibfnamefont {M.~V.}\ \bibnamefont
  {Larsen}}, \bibinfo {author} {\bibfnamefont {J.~E.}\ \bibnamefont
  {Bourassa}}, \bibinfo {author} {\bibfnamefont {S.}~\bibnamefont {Kocsis}},\
  and\ \bibinfo {author} {\bibnamefont {et~al.}},\ }\href
  {https://doi.org/10.1038/s41586-025-09044-5} {\bibfield  {journal} {\bibinfo
  {journal} {Nature}\ }\textbf {\bibinfo {volume} {642}},\ \bibinfo {pages}
  {587} (\bibinfo {year} {2025})}\BibitemShut {NoStop}%
\bibitem [{\citenamefont {Matsos}\ \emph {et~al.}(2024)\citenamefont {Matsos},
  \citenamefont {Valahu}, \citenamefont {Navickas}, \citenamefont {Rao},
  \citenamefont {Millican}, \citenamefont {Kolesnikow}, \citenamefont
  {Biercuk},\ and\ \citenamefont {Tan}}]{Matsos2024}%
  \BibitemOpen
  \bibfield  {author} {\bibinfo {author} {\bibfnamefont {V.~G.}\ \bibnamefont
  {Matsos}}, \bibinfo {author} {\bibfnamefont {C.~H.}\ \bibnamefont {Valahu}},
  \bibinfo {author} {\bibfnamefont {T.}~\bibnamefont {Navickas}}, \bibinfo
  {author} {\bibfnamefont {A.~D.}\ \bibnamefont {Rao}}, \bibinfo {author}
  {\bibfnamefont {M.~J.}\ \bibnamefont {Millican}}, \bibinfo {author}
  {\bibfnamefont {X.~C.}\ \bibnamefont {Kolesnikow}}, \bibinfo {author}
  {\bibfnamefont {M.~J.}\ \bibnamefont {Biercuk}},\ and\ \bibinfo {author}
  {\bibfnamefont {T.~R.}\ \bibnamefont {Tan}},\ }\href
  {https://doi.org/10.1103/PhysRevLett.133.050602} {\bibfield  {journal}
  {\bibinfo  {journal} {Phys. Rev. Lett.}\ }\textbf {\bibinfo {volume} {133}},\
  \bibinfo {pages} {050602} (\bibinfo {year} {2024})}\BibitemShut {NoStop}%
\bibitem [{\citenamefont {Sellem}\ \emph {et~al.}(2025)\citenamefont {Sellem},
  \citenamefont {Sarlette}, \citenamefont {Leghtas}, \citenamefont {Mirrahimi},
  \citenamefont {Rouchon},\ and\ \citenamefont {Campagne-Ibarcq}}]{Sellem2025}%
  \BibitemOpen
  \bibfield  {author} {\bibinfo {author} {\bibfnamefont {L.-A.}\ \bibnamefont
  {Sellem}}, \bibinfo {author} {\bibfnamefont {A.}~\bibnamefont {Sarlette}},
  \bibinfo {author} {\bibfnamefont {Z.}~\bibnamefont {Leghtas}}, \bibinfo
  {author} {\bibfnamefont {M.}~\bibnamefont {Mirrahimi}}, \bibinfo {author}
  {\bibfnamefont {P.}~\bibnamefont {Rouchon}},\ and\ \bibinfo {author}
  {\bibfnamefont {P.}~\bibnamefont {Campagne-Ibarcq}},\ }\href
  {https://doi.org/10.1103/PhysRevX.15.011011} {\bibfield  {journal} {\bibinfo
  {journal} {Phys. Rev. X}\ }\textbf {\bibinfo {volume} {15}},\ \bibinfo
  {pages} {011011} (\bibinfo {year} {2025})}\BibitemShut {NoStop}%
\bibitem [{\citenamefont {Connes}\ and\ \citenamefont
  {Rieffel}(1987)}]{ConnesRieffel1987}%
  \BibitemOpen
  \bibfield  {author} {\bibinfo {author} {\bibfnamefont {A.}~\bibnamefont
  {Connes}}\ and\ \bibinfo {author} {\bibfnamefont {M.~A.}\ \bibnamefont
  {Rieffel}},\ }in\ \href@noop {} {\emph {\bibinfo {booktitle} {Operator
  Algebras and Mathematical Physics (Iowa City, Iowa, 1985)}}},\ \bibinfo
  {series} {Contemp. Math.}, Vol.~\bibinfo {volume} {62}\ (\bibinfo
  {publisher} {American Mathematical Society},\ \bibinfo {address} {Providence,
  RI},\ \bibinfo {year} {1987})\ pp.\ \bibinfo {pages} {237--266}\BibitemShut
  {NoStop}%
\bibitem [{\citenamefont {Rieffel}(1988)}]{Rieffel1988}%
  \BibitemOpen
  \bibfield  {author} {\bibinfo {author} {\bibfnamefont {M.~A.}\ \bibnamefont
  {Rieffel}},\ }\href@noop {} {\bibfield  {journal} {\bibinfo  {journal} {Can.
  J. Math.}\ }\textbf {\bibinfo {volume} {40}},\ \bibinfo {pages} {257}
  (\bibinfo {year} {1988})}\BibitemShut {NoStop}%
\bibitem [{\citenamefont {Jakobsen}\ and\ \citenamefont
  {Luef}(2020)}]{Jakobsen2020}%
  \BibitemOpen
  \bibfield  {author} {\bibinfo {author} {\bibfnamefont {M.~S.}\ \bibnamefont
  {Jakobsen}}\ and\ \bibinfo {author} {\bibfnamefont {F.}~\bibnamefont
  {Luef}},\ }\href {https://doi.org/10.4171/JNCG/413} {\bibfield  {journal}
  {\bibinfo  {journal} {J. Noncommut. Geom.}\ }\textbf {\bibinfo {volume}
  {14}},\ \bibinfo {pages} {1445} (\bibinfo {year} {2020})}\BibitemShut
  {NoStop}%
\bibitem [{\citenamefont {Arnold}(1978)}]{Arnold1978}%
  \BibitemOpen
  \bibfield  {author} {\bibinfo {author} {\bibfnamefont {V.~I.}\ \bibnamefont
  {Arnold}},\ }\href@noop {} {\emph {\bibinfo {title} {Mathematical Methods of
  Classical Mechanics}}}\ (\bibinfo  {publisher} {Springer},\ \bibinfo
  {address} {New York},\ \bibinfo {year} {1978})\BibitemShut {NoStop}%
\bibitem [{\citenamefont {Joseph}\ \emph
  {et~al.}(2014{\natexlab{a}})\citenamefont {Joseph}, \citenamefont {Chew},\
  and\ \citenamefont {Sanju\'an}}]{skj_epjd}%
  \BibitemOpen
  \bibfield  {author} {\bibinfo {author} {\bibfnamefont {S.~K.}\ \bibnamefont
  {Joseph}}, \bibinfo {author} {\bibfnamefont {L.~Y.}\ \bibnamefont {Chew}},\
  and\ \bibinfo {author} {\bibfnamefont {M.~A.~F.}\ \bibnamefont {Sanju\'an}},\
  }\href@noop {} {\bibfield  {journal} {\bibinfo  {journal} {Eur. Phys. J. D}\
  }\textbf {\bibinfo {volume} {68}},\ \bibinfo {pages} {238} (\bibinfo {year}
  {2014}{\natexlab{a}})}\BibitemShut {NoStop}%
\bibitem [{\citenamefont {Joseph}\ \emph
  {et~al.}(2014{\natexlab{b}})\citenamefont {Joseph}, \citenamefont {Chew},\
  and\ \citenamefont {Sanju\'an}}]{skj_physlett}%
  \BibitemOpen
  \bibfield  {author} {\bibinfo {author} {\bibfnamefont {S.~K.}\ \bibnamefont
  {Joseph}}, \bibinfo {author} {\bibfnamefont {L.~Y.}\ \bibnamefont {Chew}},\
  and\ \bibinfo {author} {\bibfnamefont {M.~A.~F.}\ \bibnamefont {Sanju\'an}},\
  }\href@noop {} {\bibfield  {journal} {\bibinfo  {journal} {Phys. Lett. A}\
  }\textbf {\bibinfo {volume} {378}},\ \bibinfo {pages} {2603} (\bibinfo {year}
  {2014}{\natexlab{b}})}\BibitemShut {NoStop}%
\bibitem [{\citenamefont {Chung}\ and\ \citenamefont {Chew}(2007)}]{Chung07}%
  \BibitemOpen
  \bibfield  {author} {\bibinfo {author} {\bibfnamefont {N.~N.}\ \bibnamefont
  {Chung}}\ and\ \bibinfo {author} {\bibfnamefont {L.~Y.}\ \bibnamefont
  {Chew}},\ }\href@noop {} {\bibfield  {journal} {\bibinfo  {journal} {Phys.
  Rev. A}\ }\textbf {\bibinfo {volume} {76}},\ \bibinfo {pages} {032113}
  (\bibinfo {year} {2007})}\BibitemShut {NoStop}%
\bibitem [{\citenamefont {Chung}\ and\ \citenamefont
  {Chew}(2009)}]{Chung09PRA}%
  \BibitemOpen
  \bibfield  {author} {\bibinfo {author} {\bibfnamefont {N.~N.}\ \bibnamefont
  {Chung}}\ and\ \bibinfo {author} {\bibfnamefont {L.~Y.}\ \bibnamefont
  {Chew}},\ }\href@noop {} {\bibfield  {journal} {\bibinfo  {journal} {Phys.
  Rev. A}\ }\textbf {\bibinfo {volume} {80}},\ \bibinfo {pages} {012103}
  (\bibinfo {year} {2009})}\BibitemShut {NoStop}%
\bibitem [{\citenamefont {Lombardi}\ and\ \citenamefont
  {Matzkin}(2011)}]{Lombardi11}%
  \BibitemOpen
  \bibfield  {author} {\bibinfo {author} {\bibfnamefont {M.}~\bibnamefont
  {Lombardi}}\ and\ \bibinfo {author} {\bibfnamefont {A.}~\bibnamefont
  {Matzkin}},\ }\href@noop {} {\bibfield  {journal} {\bibinfo  {journal} {Phys.
  Rev. E}\ }\textbf {\bibinfo {volume} {83}},\ \bibinfo {pages} {016207}
  (\bibinfo {year} {2011})}\BibitemShut {NoStop}%
\bibitem [{\citenamefont {Lakshminarayan}(2001)}]{Arul01}%
  \BibitemOpen
  \bibfield  {author} {\bibinfo {author} {\bibfnamefont {A.}~\bibnamefont
  {Lakshminarayan}},\ }\href@noop {} {\bibfield  {journal} {\bibinfo  {journal}
  {Phys. Rev. E}\ }\textbf {\bibinfo {volume} {64}},\ \bibinfo {pages} {036207}
  (\bibinfo {year} {2001})}\BibitemShut {NoStop}%
\bibitem [{\citenamefont {Zhang}\ and\ \citenamefont {Jie}(2008)}]{Zhang08}%
  \BibitemOpen
  \bibfield  {author} {\bibinfo {author} {\bibfnamefont {S.-H.}\ \bibnamefont
  {Zhang}}\ and\ \bibinfo {author} {\bibfnamefont {Q.-L.}\ \bibnamefont
  {Jie}},\ }\href@noop {} {\bibfield  {journal} {\bibinfo  {journal} {Phys.
  Rev. A}\ }\textbf {\bibinfo {volume} {77}},\ \bibinfo {pages} {012312}
  (\bibinfo {year} {2008})}\BibitemShut {NoStop}%
\bibitem [{\citenamefont {Lombardi}\ and\ \citenamefont
  {Matzkin}(2006)}]{Lombardi06}%
  \BibitemOpen
  \bibfield  {author} {\bibinfo {author} {\bibfnamefont {M.}~\bibnamefont
  {Lombardi}}\ and\ \bibinfo {author} {\bibfnamefont {A.}~\bibnamefont
  {Matzkin}},\ }\href@noop {} {\bibfield  {journal} {\bibinfo  {journal} {Phys.
  Rev. A}\ }\textbf {\bibinfo {volume} {73}},\ \bibinfo {pages} {062335}
  (\bibinfo {year} {2006})}\BibitemShut {NoStop}%
\bibitem [{\citenamefont {Santhanam}\ \emph {et~al.}(2008)\citenamefont
  {Santhanam}, \citenamefont {Sheorey},\ and\ \citenamefont
  {Lakshminarayan}}]{Santhanam2008}%
  \BibitemOpen
  \bibfield  {author} {\bibinfo {author} {\bibfnamefont {M.~S.}\ \bibnamefont
  {Santhanam}}, \bibinfo {author} {\bibfnamefont {V.~B.}\ \bibnamefont
  {Sheorey}},\ and\ \bibinfo {author} {\bibfnamefont {A.}~\bibnamefont
  {Lakshminarayan}},\ }\href {https://doi.org/10.1103/PhysRevE.77.026213}
  {\bibfield  {journal} {\bibinfo  {journal} {Phys. Rev. E}\ }\textbf {\bibinfo
  {volume} {77}},\ \bibinfo {pages} {026213} (\bibinfo {year}
  {2008})}\BibitemShut {NoStop}%
\bibitem [{\citenamefont {Goto}\ and\ \citenamefont {Kanao}(2021)}]{Goto2021}%
  \BibitemOpen
  \bibfield  {author} {\bibinfo {author} {\bibfnamefont {H.}~\bibnamefont
  {Goto}}\ and\ \bibinfo {author} {\bibfnamefont {T.}~\bibnamefont {Kanao}},\
  }\href {https://doi.org/10.1103/PhysRevResearch.3.043196} {\bibfield
  {journal} {\bibinfo  {journal} {Phys. Rev. Res.}\ }\textbf {\bibinfo {volume}
  {3}},\ \bibinfo {pages} {043196} (\bibinfo {year} {2021})}\BibitemShut
  {NoStop}%
\bibitem [{\citenamefont {Wang}\ \emph {et~al.}(2004)\citenamefont {Wang},
  \citenamefont {Ghose}, \citenamefont {Sanders},\ and\ \citenamefont
  {Hu}}]{Wang_Ent_Indi}%
  \BibitemOpen
  \bibfield  {author} {\bibinfo {author} {\bibfnamefont {X.}~\bibnamefont
  {Wang}}, \bibinfo {author} {\bibfnamefont {S.}~\bibnamefont {Ghose}},
  \bibinfo {author} {\bibfnamefont {B.~C.}\ \bibnamefont {Sanders}},\ and\
  \bibinfo {author} {\bibfnamefont {B.}~\bibnamefont {Hu}},\ }\href@noop {}
  {\bibfield  {journal} {\bibinfo  {journal} {Phys. Rev. E}\ }\textbf {\bibinfo
  {volume} {70}},\ \bibinfo {pages} {016217} (\bibinfo {year}
  {2004})}\BibitemShut {NoStop}%
\bibitem [{\citenamefont {Bandyopadhyay}\ and\ \citenamefont
  {Lakshminarayan}(2004)}]{Bandyopadhyay04}%
  \BibitemOpen
  \bibfield  {author} {\bibinfo {author} {\bibfnamefont {J.~N.}\ \bibnamefont
  {Bandyopadhyay}}\ and\ \bibinfo {author} {\bibfnamefont {A.}~\bibnamefont
  {Lakshminarayan}},\ }\href@noop {} {\bibfield  {journal} {\bibinfo  {journal}
  {Phys. Rev. E}\ }\textbf {\bibinfo {volume} {69}},\ \bibinfo {pages} {016201}
  (\bibinfo {year} {2004})}\BibitemShut {NoStop}%
\bibitem [{\citenamefont {Chaudhury}\ \emph {et~al.}(2009)\citenamefont
  {Chaudhury}, \citenamefont {Smith}, \citenamefont {Anderson}, \citenamefont
  {Ghose},\ and\ \citenamefont {Jessen}}]{Chaudhury09}%
  \BibitemOpen
  \bibfield  {author} {\bibinfo {author} {\bibfnamefont {S.}~\bibnamefont
  {Chaudhury}}, \bibinfo {author} {\bibfnamefont {A.}~\bibnamefont {Smith}},
  \bibinfo {author} {\bibfnamefont {B.~E.}\ \bibnamefont {Anderson}}, \bibinfo
  {author} {\bibfnamefont {S.}~\bibnamefont {Ghose}},\ and\ \bibinfo {author}
  {\bibfnamefont {P.~S.}\ \bibnamefont {Jessen}},\ }\href@noop {} {\bibfield
  {journal} {\bibinfo  {journal} {Nature}\ }\textbf {\bibinfo {volume} {461}},\
  \bibinfo {pages} {768} (\bibinfo {year} {2009})}\BibitemShut {NoStop}%
\bibitem [{\citenamefont {Janssen}(1981)}]{Janssen1981}%
  \BibitemOpen
  \bibfield  {author} {\bibinfo {author} {\bibfnamefont {A.~J. E.~M.}\
  \bibnamefont {Janssen}},\ }\href
  {https://doi.org/10.1016/0022-247X(81)90130-X} {\bibfield  {journal}
  {\bibinfo  {journal} {J. Math. Anal. Appl.}\ }\textbf {\bibinfo {volume}
  {83}},\ \bibinfo {pages} {377} (\bibinfo {year} {1981})}\BibitemShut
  {NoStop}%
\bibitem [{\citenamefont {Rieffel}(1981)}]{Rieffel1981}%
  \BibitemOpen
  \bibfield  {author} {\bibinfo {author} {\bibfnamefont {M.~A.}\ \bibnamefont
  {Rieffel}},\ }\href@noop {} {\bibfield  {journal} {\bibinfo  {journal} {Pac.
  J. Math.}\ }\textbf {\bibinfo {volume} {93}},\ \bibinfo {pages} {415}
  (\bibinfo {year} {1981})}\BibitemShut {NoStop}%
\bibitem [{\citenamefont {Casazza}\ \emph {et~al.}(2001)\citenamefont
  {Casazza}, \citenamefont {Christensen},\ and\ \citenamefont
  {Janssen}}]{Janssen1999}%
  \BibitemOpen
  \bibfield  {author} {\bibinfo {author} {\bibfnamefont {P.~G.}\ \bibnamefont
  {Casazza}}, \bibinfo {author} {\bibfnamefont {O.}~\bibnamefont
  {Christensen}},\ and\ \bibinfo {author} {\bibfnamefont {A.~J. E.~M.}\
  \bibnamefont {Janssen}},\ }\href
  {https://doi.org/https://doi.org/10.1006/jfan.2000.3673} {\bibfield
  {journal} {\bibinfo  {journal} {Journal of Functional Analysis}\ }\textbf
  {\bibinfo {volume} {180}},\ \bibinfo {pages} {85} (\bibinfo {year}
  {2001})}\BibitemShut {NoStop}%
\bibitem [{\citenamefont {Connes}(1994)}]{Connes1994}%
  \BibitemOpen
  \bibfield  {author} {\bibinfo {author} {\bibfnamefont {A.}~\bibnamefont
  {Connes}},\ }\href@noop {} {\emph {\bibinfo {title} {Noncommutative
  Geometry}}}\ (\bibinfo  {publisher} {Academic Press},\ \bibinfo {address}
  {San Diego, CA},\ \bibinfo {year} {1994})\BibitemShut {NoStop}%
\bibitem [{\citenamefont {Connes}\ \emph {et~al.}(1998)\citenamefont {Connes},
  \citenamefont {Douglas},\ and\ \citenamefont {Schwarz}}]{Connes1998}%
  \BibitemOpen
  \bibfield  {author} {\bibinfo {author} {\bibfnamefont {A.}~\bibnamefont
  {Connes}}, \bibinfo {author} {\bibfnamefont {M.~R.}\ \bibnamefont
  {Douglas}},\ and\ \bibinfo {author} {\bibfnamefont {A.}~\bibnamefont
  {Schwarz}},\ }\href@noop {} {\bibfield  {journal} {\bibinfo  {journal} {J.
  High Energy Phys.}\ }\textbf {\bibinfo {volume} {02}},\bibinfo  {pages} {
  003}(\bibinfo {year} {1998})}\BibitemShut {NoStop}%
\bibitem [{\citenamefont {Manin}(2001)}]{Manin2001}%
  \BibitemOpen
\bibfield  {number} {  }\bibfield  {author} {\bibinfo {author} {\bibfnamefont
  {Y.~I.}\ \bibnamefont {Manin}},\ }\href@noop {} {\bibfield  {journal}
  {\bibinfo  {journal} {Lett. Math. Phys.}\ }\textbf {\bibinfo {volume} {56}},\
  \bibinfo {pages} {295} (\bibinfo {year} {2001})}\BibitemShut {NoStop}%
\bibitem [{\citenamefont {Manin}(2004)}]{Manin2004}%
  \BibitemOpen
  \bibfield  {author} {\bibinfo {author} {\bibfnamefont {Y.~I.}\ \bibnamefont
  {Manin}},\ }in\ \href@noop {} {\emph {\bibinfo {booktitle} {The Legacy of
  Niels Henrik Abel}}}\ (\bibinfo  {publisher} {Springer},\ \bibinfo {year}
  {2004})\ pp.\ \bibinfo {pages} {685--727}\BibitemShut {NoStop}%
\bibitem [{\citenamefont {Luef}\ and\ \citenamefont
  {Manin}(2009)}]{LuefManin2009}%
  \BibitemOpen
  \bibfield  {author} {\bibinfo {author} {\bibfnamefont {F.}~\bibnamefont
  {Luef}}\ and\ \bibinfo {author} {\bibfnamefont {Y.~I.}\ \bibnamefont
  {Manin}},\ }\href {https://doi.org/10.1007/s11005-009-0306-7} {\bibfield
  {journal} {\bibinfo  {journal} {Lett. Math. Phys.}\ }\textbf {\bibinfo
  {volume} {88}},\ \bibinfo {pages} {131} (\bibinfo {year} {2009})}\BibitemShut
  {NoStop}%
\bibitem [{\citenamefont {Casati}\ \emph {et~al.}(1979)\citenamefont {Casati},
  \citenamefont {Chirikov}, \citenamefont {Ford},\ and\ \citenamefont
  {Izrailev}}]{Casati1979}%
  \BibitemOpen
  \bibfield  {author} {\bibinfo {author} {\bibfnamefont {G.}~\bibnamefont
  {Casati}}, \bibinfo {author} {\bibfnamefont {B.~V.}\ \bibnamefont
  {Chirikov}}, \bibinfo {author} {\bibfnamefont {J.}~\bibnamefont {Ford}},\
  and\ \bibinfo {author} {\bibfnamefont {F.~M.}\ \bibnamefont {Izrailev}},\
  }in\ \href@noop {} {\emph {\bibinfo {booktitle} {Lecture Notes in
  Physics}}},\ Vol.~\bibinfo {volume} {93}\ (\bibinfo  {publisher} {Springer},\
  \bibinfo {year} {1979})\ p.\ \bibinfo {pages} {334}\BibitemShut {NoStop}%
\bibitem [{\citenamefont {Izrailev}(1990)}]{Izrailev1990}%
  \BibitemOpen
  \bibfield  {author} {\bibinfo {author} {\bibfnamefont {F.~M.}\ \bibnamefont
  {Izrailev}},\ }\href {https://doi.org/10.1016/0370-1573(90)90067-C}
  {\bibfield  {journal} {\bibinfo  {journal} {Phys. Rep.}\ }\textbf {\bibinfo
  {volume} {196}},\ \bibinfo {pages} {299} (\bibinfo {year}
  {1990})}\BibitemShut {NoStop}%
\bibitem [{\citenamefont {Bhosale}\ and\ \citenamefont
  {Santhanam}(2017)}]{Santhanam2017}%
  \BibitemOpen
  \bibfield  {author} {\bibinfo {author} {\bibfnamefont {U.~T.}\ \bibnamefont
  {Bhosale}}\ and\ \bibinfo {author} {\bibfnamefont {M.~S.}\ \bibnamefont
  {Santhanam}},\ }\href {https://doi.org/10.1103/PhysRevE.95.012216} {\bibfield
   {journal} {\bibinfo  {journal} {Phys. Rev. E}\ }\textbf {\bibinfo {volume}
  {95}},\ \bibinfo {pages} {012216} (\bibinfo {year} {2017})}\BibitemShut
  {NoStop}%
\bibitem [{\citenamefont {Hannay}\ and\ \citenamefont
  {Berry}(1980)}]{Hannay1980}%
  \BibitemOpen
  \bibfield  {author} {\bibinfo {author} {\bibfnamefont {J.~H.}\ \bibnamefont
  {Hannay}}\ and\ \bibinfo {author} {\bibfnamefont {M.~V.}\ \bibnamefont
  {Berry}},\ }\href {https://doi.org/10.1016/0167-2789(80)90026-3} {\bibfield
  {journal} {\bibinfo  {journal} {Phys. D}\ }\textbf {\bibinfo {volume} {1}},\
  \bibinfo {pages} {267} (\bibinfo {year} {1980})}\BibitemShut {NoStop}%
\bibitem [{\citenamefont {Moore}(2017)}]{Moore2017}%
  \BibitemOpen
  \bibfield  {author} {\bibinfo {author} {\bibfnamefont {G.~W.}\ \bibnamefont
  {Moore}},\ }\href {https://arxiv.org/abs/1701.07746} {\bibinfo {title}
  {Quantum mechanics with noncommutative amplitudes}} (\bibinfo {year}
  {2017}),\ \Eprint {https://arxiv.org/abs/1701.07746} {arXiv:1701.07746
  [hep-th]} \BibitemShut {NoStop}%
\bibitem [{\citenamefont {Groenewold}(1946)}]{Groenewold1946}%
  \BibitemOpen
  \bibfield  {author} {\bibinfo {author} {\bibfnamefont {H.~J.}\ \bibnamefont
  {Groenewold}},\ }\href {https://doi.org/10.1016/S0031-8914(46)80059-4}
  {\bibfield  {journal} {\bibinfo  {journal} {Physica}\ }\textbf {\bibinfo
  {volume} {12}},\ \bibinfo {pages} {405} (\bibinfo {year} {1946})}\BibitemShut
  {NoStop}%
\bibitem [{\citenamefont {Moyal}(1949)}]{Moyal1949}%
  \BibitemOpen
  \bibfield  {author} {\bibinfo {author} {\bibfnamefont {J.~E.}\ \bibnamefont
  {Moyal}},\ }\href {https://doi.org/10.1017/S0305004100000487} {\bibfield
  {journal} {\bibinfo  {journal} {Proc. Cambridge Phil. Soc.}\ }\textbf
  {\bibinfo {volume} {45}},\ \bibinfo {pages} {99} (\bibinfo {year}
  {1949})}\BibitemShut {NoStop}%
\bibitem [{\citenamefont {D'Ascanio}\ \emph {et~al.}(2016)\citenamefont
  {D'Ascanio}, \citenamefont {Pisani},\ and\ \citenamefont
  {Vassilevich}}]{DAscanio2016}%
  \BibitemOpen
  \bibfield  {author} {\bibinfo {author} {\bibfnamefont {D.}~\bibnamefont
  {D'Ascanio}}, \bibinfo {author} {\bibfnamefont {P.}~\bibnamefont {Pisani}},\
  and\ \bibinfo {author} {\bibfnamefont {D.~V.}\ \bibnamefont {Vassilevich}},\
  }\href {https://doi.org/10.1140/epjc/s10052-016-4022-z} {\bibfield  {journal}
  {\bibinfo  {journal} {Eur. Phys. J. C}\ }\textbf {\bibinfo {volume} {76}},\
  \bibinfo {pages} {180} (\bibinfo {year} {2016})}\BibitemShut {NoStop}%
\bibitem [{\citenamefont {Zak}(1964)}]{Zak1964}%
  \BibitemOpen
  \bibfield  {author} {\bibinfo {author} {\bibfnamefont {J.}~\bibnamefont
  {Zak}},\ }\href {https://doi.org/10.1103/PhysRev.134.A1602} {\bibfield
  {journal} {\bibinfo  {journal} {Phys. Rev.}\ }\textbf {\bibinfo {volume}
  {134}},\ \bibinfo {pages} {A1602} (\bibinfo {year} {1964})}\BibitemShut
  {NoStop}%
\bibitem [{\citenamefont {Pantaleoni}\ \emph {et~al.}(2023)\citenamefont
  {Pantaleoni}, \citenamefont {Baragiola},\ and\ \citenamefont
  {Menicucci}}]{Pantaleoni2023}%
  \BibitemOpen
  \bibfield  {author} {\bibinfo {author} {\bibfnamefont {G.}~\bibnamefont
  {Pantaleoni}}, \bibinfo {author} {\bibfnamefont {B.~Q.}\ \bibnamefont
  {Baragiola}},\ and\ \bibinfo {author} {\bibfnamefont {N.~C.}\ \bibnamefont
  {Menicucci}},\ }\href {https://doi.org/10.1103/PhysRevA.107.062611}
  {\bibfield  {journal} {\bibinfo  {journal} {Phys. Rev. A}\ }\textbf {\bibinfo
  {volume} {107}},\ \bibinfo {pages} {062611} (\bibinfo {year}
  {2023})}\BibitemShut {NoStop}%
\bibitem [{\citenamefont {Zayed}\ and\ \citenamefont
  {Mikusi{\'n}ski}(1995)}]{Zayed1995}%
  \BibitemOpen
  \bibfield  {author} {\bibinfo {author} {\bibfnamefont {A.~I.}\ \bibnamefont
  {Zayed}}\ and\ \bibinfo {author} {\bibfnamefont {P.}~\bibnamefont
  {Mikusi{\'n}ski}},\ }\href {https://doi.org/10.4310/MAA.1995.v2.n2.a3}
  {\bibfield  {journal} {\bibinfo  {journal} {Methods Appl. Anal.}\ }\textbf
  {\bibinfo {volume} {2}},\ \bibinfo {pages} {160} (\bibinfo {year}
  {1995})}\BibitemShut {NoStop}%
\bibitem [{\citenamefont {Rai}\ and\ \citenamefont
  {Mehta}(1988)}]{JagdishRai88}%
  \BibitemOpen
  \bibfield  {author} {\bibinfo {author} {\bibfnamefont {J.}~\bibnamefont
  {Rai}}\ and\ \bibinfo {author} {\bibfnamefont {C.~L.}\ \bibnamefont
  {Mehta}},\ }\href@noop {} {\bibfield  {journal} {\bibinfo  {journal} {Phys.
  Rev. A}\ }\textbf {\bibinfo {volume} {37}},\ \bibinfo {pages} {4497}
  (\bibinfo {year} {1988})}\BibitemShut {NoStop}%
\bibitem [{\citenamefont {M{\o}ller}\ \emph {et~al.}(1996)\citenamefont
  {M{\o}ller}, \citenamefont {J{\o}rgensen},\ and\ \citenamefont
  {Dahl}}]{Moller96}%
  \BibitemOpen
  \bibfield  {author} {\bibinfo {author} {\bibfnamefont {K.~B.}\ \bibnamefont
  {M{\o}ller}}, \bibinfo {author} {\bibfnamefont {T.~G.}\ \bibnamefont
  {J{\o}rgensen}},\ and\ \bibinfo {author} {\bibfnamefont {J.~P.}\ \bibnamefont
  {Dahl}},\ }\href@noop {} {\bibfield  {journal} {\bibinfo  {journal} {Phys.
  Rev. A}\ }\textbf {\bibinfo {volume} {54}},\ \bibinfo {pages} {5378}
  (\bibinfo {year} {1996})}\BibitemShut {NoStop}%
\bibitem [{\citenamefont {Mumford}(1983)}]{MumfordBook1}%
  \BibitemOpen
  \bibfield  {author} {\bibinfo {author} {\bibfnamefont {D.}~\bibnamefont
  {Mumford}},\ }\href@noop {} {\emph {\bibinfo {title} {Tata Lectures on Theta
  I}}}\ (\bibinfo  {publisher} {Birkh{\"a}user},\ \bibinfo {address} {Boston},\
  \bibinfo {year} {1983})\BibitemShut {NoStop}%
\bibitem [{\citenamefont {Deconinck}\ \emph {et~al.}(2004)\citenamefont
  {Deconinck}, \citenamefont {Heil}, \citenamefont {Bobenko}, \citenamefont
  {van Hoeij},\ and\ \citenamefont {Schmies}}]{Bernard2004}%
  \BibitemOpen
  \bibfield  {author} {\bibinfo {author} {\bibfnamefont {B.}~\bibnamefont
  {Deconinck}}, \bibinfo {author} {\bibfnamefont {M.}~\bibnamefont {Heil}},
  \bibinfo {author} {\bibfnamefont {A.~I.}\ \bibnamefont {Bobenko}}, \bibinfo
  {author} {\bibfnamefont {M.}~\bibnamefont {van Hoeij}},\ and\ \bibinfo
  {author} {\bibfnamefont {M.}~\bibnamefont {Schmies}},\ }\href
  {https://doi.org/10.1090/S0025-5718-03-01609-0} {\bibfield  {journal}
  {\bibinfo  {journal} {Math. Comput.}\ }\textbf {\bibinfo {volume} {73}},\
  \bibinfo {pages} {1417} (\bibinfo {year} {2004})}\BibitemShut {NoStop}%
\bibitem [{\citenamefont {Abramowitz}\ and\ \citenamefont
  {Stegun}(1970)}]{Abramowitz1970}%
  \BibitemOpen
  \bibinfo {editor} {\bibfnamefont {M.}~\bibnamefont {Abramowitz}}\ and\
  \bibinfo {editor} {\bibfnamefont {I.~A.}\ \bibnamefont {Stegun}},\ eds.,\
  \href@noop {} {\emph {\bibinfo {title} {Handbook of Mathematical Functions
  with Formulas, Graphs, and Mathematical Tables}}},\ \bibinfo {series}
  {Applied Mathematics Series}\ No.~\bibinfo {number} {55}\ (\bibinfo
  {publisher} {U.S. Department of Commerce, National Bureau of Standards},\
  \bibinfo {address} {Washington, D.C.},\ \bibinfo {year} {1970})\BibitemShut
  {NoStop}%
\bibitem [{\citenamefont {Kowalski}\ \emph {et~al.}(1996)\citenamefont
  {Kowalski}, \citenamefont {Rembieli{\'n}ski},\ and\ \citenamefont
  {Papaloucas}}]{Kowalski1996}%
  \BibitemOpen
  \bibfield  {author} {\bibinfo {author} {\bibfnamefont {K.}~\bibnamefont
  {Kowalski}}, \bibinfo {author} {\bibfnamefont {J.}~\bibnamefont
  {Rembieli{\'n}ski}},\ and\ \bibinfo {author} {\bibfnamefont {L.~C.}\
  \bibnamefont {Papaloucas}},\ }\href
  {https://doi.org/10.1088/0305-4470/29/14/034} {\bibfield  {journal} {\bibinfo
   {journal} {J. Phys. A: Math. Theor.}\ }\textbf {\bibinfo {volume} {29}},\
  \bibinfo {pages} {4149} (\bibinfo {year} {1996})}\BibitemShut {NoStop}%
\bibitem [{\citenamefont {Kowalski}\ and\ \citenamefont
  {Ławniczak}(2021)}]{Kowalski2021}%
  \BibitemOpen
  \bibfield  {author} {\bibinfo {author} {\bibfnamefont {K.}~\bibnamefont
  {Kowalski}}\ and\ \bibinfo {author} {\bibfnamefont {K.}~\bibnamefont
  {Ławniczak}},\ }\href {https://doi.org/10.1088/1751-8121/ac019d} {\bibfield
  {journal} {\bibinfo  {journal} {J. Phys. A: Math. Theor.}\ }\textbf {\bibinfo
  {volume} {54}},\ \bibinfo {pages} {275302} (\bibinfo {year}
  {2021})}\BibitemShut {NoStop}%
\end{thebibliography}

\end{document}